\documentclass[superscriptaddress,twocolumn,secnumarabic,
amssymb,amsmath,nobibnotes,aps,prd,showkeys,showpacs,nofootinbib]{revtex4}%
\usepackage{graphicx}
\usepackage{epsf}
\usepackage{bm}
\usepackage{amsmath}
\usepackage{amsfonts}
\usepackage{amssymb}%
\providecommand{\U}[1]{\protect\rule{.1in}{.1in}}
\newcommand{\be}{\begin{equation}}
\newcommand{\ee}{\end{equation}}

\newcommand{\mincir}{\raise
-3.truept\hbox{\rlap{\hbox{$\sim$}}\raise4.truept\hbox{$<$}\ }}
\newcommand{\magcir}{\raise
-3.truept\hbox{\rlap{\hbox{$\sim$}}\raise4.truept\hbox{$>$}\ }}

\begin{document}
\title{Hyperbolic Inflation in the Light of \emph{Planck 2015} data}
\author{Spyros Basilakos}
\affiliation{Academy of Athens, Research Center for Astronomy and Applied Mathematics,
Soranou Efessiou 4, 115 27 Athens, Greece. }
\author{John D. Barrow}
\affiliation{DAMTP, Centre for Mathematical Sciences, University of Cambridge, Wilberforce
Rd., Cambridge CB3 0WA, U.K.}

\begin{abstract}
\vspace{0.2cm}

Rubano and Barrow have discussed the emergence of a dark energy, with
late-time cosmic acceleration arising from a self-interacting homogeneous
scalar field with a potential of hyperbolic power type. Here, we study the
evolution of this scalar field potential back in the inflationary era. Using
the hyperbolic power potential in the framework of inflation, we find that the
main slow-roll parameters, like the scalar spectral index, the running of the
spectral index and the tensor-to-scalar fluctuation ratio can be computed
analytically. Finally, in order to test the viability of this hyperbolic
scalar field model at the early stages of the Universe, we compare the
predictions of that model against the latest observational data, namely
\textit{Planck 2015}.

\end{abstract}

\maketitle


\hyphenation{tho-rou-ghly in-te-gra-ting e-vol-ving con-si-de-ring
ta-king me-tho-do-lo-gy fi-gu-re}

\section{Introduction \label{sec:1}}

The study of Cosmic Microwave Background (CMB) photons using the
\textit{Planck} \cite{Planck} and BICEP2 \cite{Bicep2} data sets has opened up
a new constraint on inflationary models~\cite{encyclo}. Specifically, the
detailed analysis of \textit{Planck} data~\cite{Planck} constrains single
scalar-field models of slow-roll inflation, to have very low tensor-to-scalar
fluctuation ratio $r=n_{t}/n_{s}\ll1$, with a scalar spectral index
$n_{s}\simeq0.96$ possessing no appreciable running. The upper bound set by
the \textit{Planck} Collaboration~\cite{Planck} on tensor-to-scalar
fluctuation ratio, related to the absence of the B-modes of polarization in
the CMB, is $r<0.11$, but their favored regions (higher than 95\% C.L.) point
towards a more stringent bound of $r\leq10^{-3}$. From the theoretical view
point one can see that this is in agreement with the so-called
Starobinsky-type (or $R^{2}$, with $R$ denoting the scalar space-time
curvature) inflationary models~\cite{staro}.

Last year, however, the BICEP2 team~\cite{Bicep2} made an important
claim: to have made the first measurement of B-mode polarization in the CMB
radiation. This measurement was initially interpreted as indication
for gravitational waves at the time of the last scattering, with a
tensor-to-scalar ratio $r=0.16_{-0.05}^{+0.06}$.
If such claims were confirmed, they would constitute the first experimental
observation of (transverse) primordial (possibly quantum) metric fluctuations.
For the scalar spectal index, it has been found that $n_{s}\simeq0.96$
and $dn_{s}/d\mathrm{ln}k\simeq0$, in agreement with the \textit{Planck}
data~\cite{Planck}.

Since then a great effort has been spent in order to compare the BICEP2
tensor-to-scalar ratio against the choice of inflationary paradigm ($R^{2}$
\cite{staro}, chaotic \cite{Linde}, inverse power law \cite{Barr}, hilltop
\cite{Alb}, natural \cite{Free}, supersymmetry \cite{Mav}, D-flation
\cite{Mav1} and the like).

Furthermore, during this period there was an intense debate as to whether the
BICEP2 signal is indubitably due to primordial gravitational waves, or is
polluted by gravitationally-lensed E-modes and Galactic foregrounds.
Recently,
in~ref. \cite{Subir} it was stressed
that magnetized dust associated with radio loops due to supernova remnants
might affect the signal received by BICEP2.
In the case of polarization effects by galactic dust, refs. ~\cite{MS}
and~\cite{Flauger} have shown that the cosmological value of tensor-to-scalar
ratio could be very small $r\ll0.1$. Within this framework, the recent
analysis on the foreground dust in the BICEP2 region released by the
\textit{Planck} collaboration~\cite{Planckrec} points to a significant
foreground pollution which means that the BICEP2 B-mode polarization data
cannot be used as evidence for primordial CMB polarization.

Recently, the joint analysis of BICEP2/Keck Array and \textit{Planck} data
appeared in the literature \cite{joint}. This analysis has placed an upper
bound in the tensor-to-scalar ratio, namely $r<0.12$ at $95\%$ significance
level. In this context, the Planck team has repeated the inflationary analysis
by using the \textit{Planck 2015} data and essentially confirm the
\textit{Planck 2013} results: $n_{s}=0.968\pm0.006$, $dn_{s}/d\mathrm{ln}%
k=-0.003\pm0.007$ and $r<0.11$ \cite{Planck2015}.


The crux of these studies is that the potential energy of the scalar field, is
not really known and one must introduce it using phenomenological arguments,
starting with the simplest possibilities. There has been intense debate and
speculation about the functional form of the potential energy $V(\phi)$.
Various candidates have been proposed in the literature, such as a power law,
inverse power law, exponential and so on (for review see \cite{Lyth} and
references therein) but exact solutions are not abundant and always possess
special mathematical features.

Some time ago, a special solution for a spatially flat
Friedmann-Lema\^{\i}tre-Robertson-Walker (FLRW) spacetime with a perfect fluid
with a constant equation of state parameter\textbf{ }$P_{m}=(\gamma-1)\rho
_{m}$\textbf{ }(where, for radiation $\gamma=4/3$ and for dust $\gamma=1$) and
a scalar field with a constant equation of state
$P_{\phi}=(\gamma_{\phi}-1)\rho_{\phi}$ was found in \cite{Barrow}. In
particular, Rubano and Barrow \cite{Barrow} (see also \cite{Urena} and
\cite{Saa}) showed that under of specific conditions we can solve the Einstein
equations when the potential $V(\phi)$ has the interesting hyperbolic form:
\begin{equation}
V\left(  \phi\right)  =A\left[  \sinh\left(  \sqrt{3}\frac{(\gamma
-\gamma_{\phi})}{\sqrt{\gamma_{\phi}}}\left(  \phi-\phi_{0}\right)  \right)
\right]  ^{b}, \label{pot}%
\end{equation}
where the constant $A$ is
\begin{equation}
A=3H_{0}^{2}\left(  1-\Omega_{m0}\right)  \left(  1-\frac{\gamma_{\phi}}%
{2}\right)  \left(  \frac{1-\Omega_{m0}}{\Omega_{m0}}\right)  ^{-b/2}
\label{AA}%
\end{equation}
and
\begin{equation}
b=-\frac{2\gamma_{\phi}}{\gamma-\gamma_{\phi}}=\frac{2(1+w_{\phi})}{1+w_{\phi
}-\gamma}. \label{bb}%
\end{equation}
Note that $H_{0}$ and $\Omega_{m0}$ are the usual cosmological parameters
(although $\Omega_{m0}$ denotes any possible matter content according to the
appropriate choice of $\gamma$). If we trace the late universe (dustlike
matter $P_{m}=0$) we have $\gamma=1$. 
We remind the reader that $w_{\phi}$ denotes the equation of state parameter
of the dark energy usually parametrized by $w_{\phi}=\frac{P_{\phi}}%
{\rho_{\phi}}=\gamma_{\phi}-1$, with $P_{\phi}$ and $\rho_{\phi}$ being the
pressure and density of the dark energy fluid.
Obviously the exponent $b$ plays an important role in the cosmic dynamics,
since it is related with the equation of state parameter, $w_{\phi}$. In this
context, inverting Eq.(\ref{bb}) one can prove
that the equation of state parameter reduces to $w_{\phi}=2/(b-2)$. Note that
the accelerated expansion of the universe poses the restriction $w_{\phi
}<-1/3\Omega_{de0}$ which implies $2(1-3\Omega_{de0})<b<2$, where
$\Omega_{de0}=1-\Omega_{m0}$.

The potential (\ref{pot}) has some interesting geometric characteristics.
Under specific conditions, it behaves either as an exponential or as a
power-law. If $|\lambda\phi|\gg1$ (or $|\lambda\phi|\ll1$),
we find $V\propto e^{-b\lambda\phi}$ (or $V\propto(\lambda\phi)^{b}$), where
$\lambda=\frac{\sqrt{3}(1-\gamma_{\phi})}{\sqrt{\gamma_{\phi}}}$ (see also
\cite{Saa}). The initial motivation in \cite{Barrow} was to use potential
(\ref{pot}) to describe the late-time acceleration of the universe, but we can
also apply Eq.(\ref{pot}) to the very early states of the cosmic evolution
when the cosmic fluid is dominated by the radiation and the inflaton
components.\textbf{ }This might then provide a unified picture of inflation
and dark energy in which both eras are described by the potential of
Eq.(\ref{pot}). It is the purpose of this work to demonstrate compatibility of
the Rubano and Barrow \cite{Barrow} scenario with the \textit{Planck 2015}
data,
taking into account the foreground ambiguities clouding the \textit{Planck
2015} data, as mentioned previously. The structure of the paper is as follows.
The slow-roll inflation and its connection to the hyperbolic potential of ref.
\cite{Barrow} are reviewed in section II. In section III we study the
performance of Eq.(\ref{pot}) against the \textit{Planck 2015} data.
Finally, our conclusions are summarized in section IV.

\begin{figure}[ptb]
\includegraphics[width=0.48\textwidth]{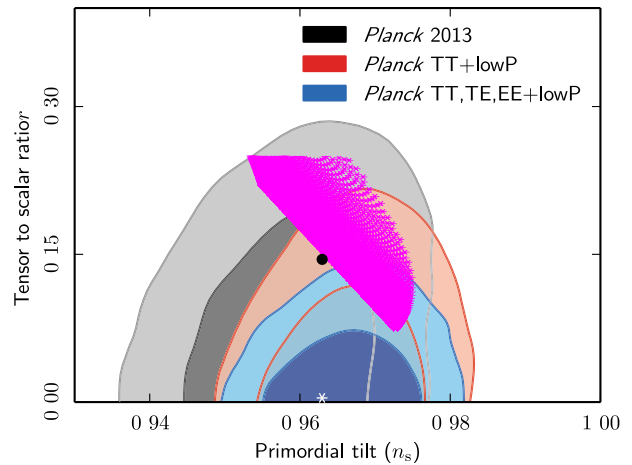}\caption{The $n_{s}-r$ diagram for
the hyperbolic potential of \cite{Saa} using $N=55$. The contours borrowed
from \emph{Planck 2015} \cite{Planck2015}. The area which is plotted over the
contours corresponds to the hyperbolic Rubano \& Barrow \cite{Barrow}
potential (see Eq.(\ref{pot1}). The points correspond to chaotic (solid point)
and Starobinsky (star) inflation respectively.}%
\label{fig:1}%
\end{figure}

\begin{figure}[ptb]
\caption{The $n_{s}-n_{s}^{\prime}$ diagram. The area corresponds to the
hyperbolic Rubano \& Barrow \cite{Barrow} potential (for more details see the
caption of figure 1).}%
\label{fig:2}%
\includegraphics[width=0.48\textwidth]{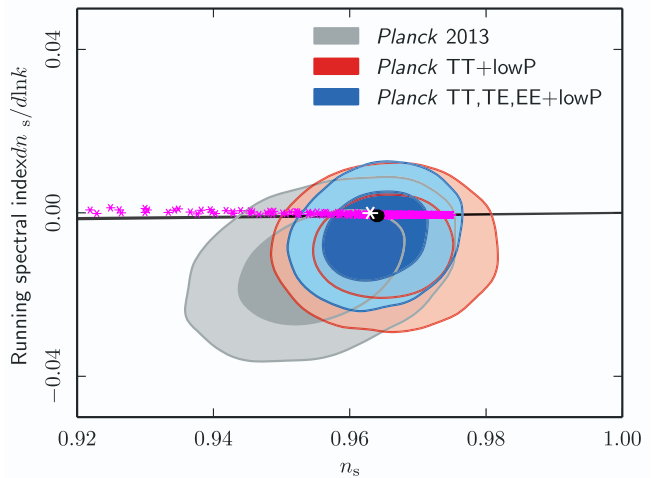}\end{figure}

\section{Slow-roll inflation}

Let us present here the basic ingredients in the context of single-field
inflation. Assuming an inflaton field $\phi$ with potential energy $V(\phi)$,
the slow-roll parameters are given by
\begin{equation}
\epsilon=\frac{M_{pl}^{2}V^{^{\prime}2}}{2V^{2}}, \label{ee}%
\end{equation}%
\begin{equation}
\eta=\frac{M_{pl}^{2}V^{^{\prime\prime}}}{V} \label{nn}%
\end{equation}%
\begin{equation}
\xi=\frac{M_{pl}^{4}V^{^{\prime}}V^{^{\prime\prime\prime}}}{V^{2}}, \label{xx}%
\end{equation}
where the prime denotes derivatives with respect to $\phi$ and $M_{pl}%
^{2}=1/8\pi G$. The corresponding spectral indices are defined in terms of the
slow-roll parameters, as usual \cite{LL}, by
\begin{equation}
n_{s}\simeq1+2\eta-6\epsilon\label{sp1}%
\end{equation}%
\begin{equation}
r\simeq16\epsilon\simeq-8n_{t} \label{sp2}%
\end{equation}%
\begin{equation}
n_{s}^{^{\prime}}=dn_{s}/d\mathrm{ln}k\simeq16\epsilon\eta-24\epsilon^{2}%
-2\xi. \label{sp3}%
\end{equation}
In this framework, the number of e-folds is written as\footnote{In the
literature sometimes we label $\phi$ by $\phi_{\star}$ which denotes the value
at the horizon crossing for which the scalar amplitude is $A_{s}\approx
\Lambda^{4}/24\pi^{2}\epsilon M_{pl}^{4}$, where the energy scale of inflation
is $\Lambda\sim10^{16}$Gev.}
\begin{equation}
N=\int_{t}^{t_{end}}H(t)dt\simeq\frac{1}{M_{pl}^{2}}\int_{\phi_{end}}^{\phi
}\frac{V(\phi)}{V^{^{\prime}}(\phi)}d\phi\;, \label{efold}%
\end{equation}
where $\phi_{end}$ is the value of the inflaton field at the end of inflation,
namely $\epsilon(\phi_{end})\simeq1$. Now, let us focus on the potential of
Eq.(\ref{pot}) which is written as
\begin{equation}
V(\phi)=A\;\mathrm{sinh}^{b}(\phi/f), \label{pot1}%
\end{equation}
where $f$ is the scale in units of $M_{pl}$.
Recall that in the dark-energy era the corresponding constants are related
with the cosmological parameters [see Eqs.(\ref{AA}) and (\ref{bb})].
Therefore, in the early universe one may expect that the corresponding
constants in Eq.(\ref{pot1}) are not necessarily equal to those derived by
\cite{Barrow} using arguments from the late universe where pressureless matter
dominates radiation (for comparison with the observational data see
\cite{Saa}). When radiation dominates we have $\gamma=4/3$ and so from the
second equality of Eq.(\ref{bb}) we arrive at
\begin{equation}
\label{bbw}b=\frac{6(1+w_{\phi})}{3w_{\phi}-1}.
\end{equation}
Note that the restriction $b>1$ implies that
the potential contains a critical point.

Now inserting Eq.(\ref{pot1}) into the slow-roll parameters we obtain after
some simple algebra
\begin{equation}
\epsilon=\frac{b^{2}M_{pl}^{2}}{2f^{2}}\mathrm{coth}^{2}(\phi/f), \label{ee1}%
\end{equation}%
\begin{equation}
\eta=\frac{bM_{pl}^{2}}{f^{2}}\left[  (b-1)\mathrm{coth}^{2}(\phi/f)+1\right]
, \label{nn1}%
\end{equation}%
\begin{equation}
\xi=\frac{b^{2}M_{pl}^{4}}{f^{4}}\mathrm{coth}^{2}(\phi/f)\left[
(b-1)(b-2)\mathrm{coth}^{2}(\phi/f)+(3b-2)\right]  . \label{xx1}%
\end{equation}
Combining Eq.(\ref{ee1}) with Eq.(\ref{nn1}) and Eq.(\ref{xx1}) we can write
$(\eta,\xi)$ in terms of $\epsilon$:
\begin{equation}
\eta=\frac{bM_{pl}^{2}}{f^{2}}\left[  \frac{2f^{2}\epsilon}{b^{2}M_{pl}^{2}%
}(b-1)+1\right]  \label{nn2}%
\end{equation}

\begin{equation}
\label{xx2}\xi=\frac{2M_{pl}^{2} \epsilon}{f^{2}} \left[  \frac{2f^{2}%
(b-1)(b-2)\epsilon}{b^{2}M_{pl}^{2}}+(3b-2) \right]  .
\end{equation}

In this case, the number of e-folds becomes
\begin{equation}
N\simeq\frac{f^{2}}{bM_{pl}^{2}}\mathrm{ln}\left[  \frac{\mathrm{cosh}%
(\phi/f)}{\mathrm{cosh}(\phi_{end}/f)}\right]  . \label{efold1}%
\end{equation}
Inverting the above, we can express the inflaton as $\phi(N)$ by
\begin{equation}
\phi=f\;\mathrm{cosh}^{-1}\left[  e^{NbM_{pl}^{2}/f^{2}}\mathrm{cosh}%
(\phi_{end}/f)\right]  . \label{efold2}%
\end{equation}
In order to proceed with the analysis we need to know the values of $N$ and
$\phi_{end}$. Firstly, it is natural to consider that the number of e-folds
lies in the interval $[50,60]$. Here, we set it to 55, for concreteness.
Secondly, using the constraint $\epsilon(\phi_{end})\simeq1$ and
Eq.(\ref{ee1}) we can estimate the value of the scalar field at the end of
inflation to be
\begin{equation}
\phi_{end}\simeq\frac{f}{2}\mathrm{ln}\left(  \frac{\theta+1}{\theta
-1}\right)  ~, \label{efold11}%
\end{equation}
where $\theta=\sqrt{2}f/bM_{pl}$. Since $\theta>1,$ the scale $f$ obeys the
restriction $f>\sqrt{2}bM_{pl}/2$.


\section{Observational restrictions}

The point of this section is to test the viability of the potential
(\ref{pot1}) at the inflationary level, involving the latest cosmological
data. In particular, the combined analysis between \emph{Planck 2015} and
various data such as WMAP, high-$l$ data, and Baryonic Acoustic Oscillations
(BAO), shows that the scalar spectral index is $n_{s}=0.968\pm0.006$. For the
scalar spectral index in this work we use $n_{s}^{\prime}=-0.003\pm0.007$.
Furthermore, as mentioned in the introduction, the analysis of the
\emph{Planck} collaboration places an upper limit on the tensor-to-scalar
ratio is concerned, $r<0.11,$ which is in agreement with the joint analysis of
BICEP2/Keck Array and \textit{Planck} \cite{joint}.

Let us now briefly present our results. In figures 1 and 2 we show the
confidence contours in the $(n_{s},r)$ and $(n_{s},n_{s}^{\prime})$ planes
which are provided by the Planck team \cite{Planck2015}. On top of that we
present the area for the individual sets of $(n_{s},r)$ which are based on the
potential (\ref{pot1}), whereas in figure 2 we display the corresponding area
in the case of $(n_{s},n_{s}^{\prime})$. Obviously, our results are consistent
with those of \textit{Planck 2015}. Specifically, as it can be seen from
figure 1, the tensor-to-scalar fluctuation ratio could reach the value of
$r\simeq0.075$ which is in a good agreement within $1\sigma$ with that of
BICEP2/Keck Array/\textit{Planck} results ($r\simeq0.05$ see figure 9 in
\cite{joint}). Notice that in order to derive $r\simeq0.07$ the constants in
Eq.(\ref{pot1}) need to obey the following inequalities: $1.02\leq b\leq1.1$
[or inverting Eq.(\ref{bbw}) we obtain $-2.63 \le w_{\phi} \le-2.39$]
and $26M_{pl}\leq f\leq39M_{pl}$. Concerning the running spectral index we
obtain $n_{s}^{\prime}\simeq-0.004$ which is consistent with that of
\emph{Planck 2015}, namely $n_{s}^{\prime}=-0.003\pm0.007$. Moreover, we find
that the running of the scalar spectral index does not change significantly as
a function of $n_{s}$ (see figure 2). Regarding the inflaton field at the
beginning of inflation, we obtain $10.6M_{pl}\leq\phi\leq12.9M_{pl}$; while at
the end of inflation we require $0.7M_{pl}\leq\phi_{end}\leq1.1M_{pl}$.
Furthermore, we find that the allowed region in which our results satisfy the
$2\sigma$ observational restrictions of \emph{Planck 2015} is $f\geq
11.7M_{pl}$ and $1<b\leq1.5$.
Thus the data applied to slow-roll inflation place constraints on $b$,
although the value of scale $f$ has only a lower limit at $2\sigma$ level.


Finally, we would like to compare our results with those found using other
potentials. Specifically, in the case of the chaotic inflation $V(\phi
)=\Lambda^{4}(\phi/M_{pl})^{k}$ \cite{Linde}, the corresponding slow-roll
parameters are written as $\epsilon=k/4N$, $\eta=(k-1)/2N$ and $\xi
=(k-1)(k-2)/4N^{2}$. The latter implies $n_{s}=1-(k+2)/2N$, $r=4k/N$ and
$n_{s}^{\prime}=-(2+k)/2N^{2}$. Using $k=2$ and $N=55$ we obtain $n_{s}%
\simeq0.964$, $r\simeq0.145$ and $n_{s}^{\prime}\simeq-0.0007$. This also
corresponds to the slow-roll regime of intermediate inflation \cite{int} with
Hubble rate during inflation given by $H\propto t^{k/(4-k)\text{ }}$with
$n_{s}=1-(k+2)r/8k$ and $k=-2$ gives $n_{s}=1$ exactly to first order and
$n_{s}^{^{\prime}}=-2(n_{s}-1)^{2}/(k+2)$. On the other hand the Starobinsky
inflation \cite{staro}, namely $V(\phi)\propto\lbrack1-2\mathrm{e}%
^{-B\phi/M_{pl}}+\mathcal{O}(\mathrm{e}^{-2B\phi/M_{pl}})]$
, leads to the following slow-roll predictions \cite{Muk81,Ellis13}:
$n_{s}\approx1-2/N$ and $r\approx8/B^{2}N^{2}$, where $B^{2}=2/3$.
Furthermore, for Starobinsky inflation following the notations of
\cite{Ellis13} we find that the running spectral index is given by
$n_{s}^{\prime}\approx-2/N^{2}$. To this end using $N=55$ we obtain
$(n_{s},r,n_{s}^{\prime})\approx(0.963,0.004,-6.6\times10^{-4})$. The above
slow-roll parameters are indicated by the solid points (chaotic inflation) in
figures 1 and 2 while the stars represent values for the Starobinsky
inflation. \emph{\ }

\section{Conclusions}

In the light of the \emph{Planck 2015} results, a debate is taking place in
the literature about the best implementation of the inflationary paradigm. In
the current article we would like to contribute to this discussion. Employing
the hyperbolic dark-energy scalar-field potential of Rubano and Barrow
\cite{Barrow} (see also \cite{Saa}) we study the performance of this model as
a description of inflation. We find that the hyperbolic inflation turns out to
be quite promising in the context of the new data from \emph{Planck 2015}.
Specifically, using the scalar-field potential (\ref{pot1}), we calculate the
slow-roll parameters analytically and then we compare the corresponding
predictions against the observational data. We find that currently hyperbolic
inflation is consistent with the results provided by \emph{Planck 2015} within
$1\sigma$ uncertainties. The combination of ref \cite{Barrow} with
these calculations provide an overall
cosmological investigation of the potential given by Eq.(\ref{pot1}). We find
that the hyperbolic structure of this potential leads to a viable model which
can be used separately to understand the main properties of both inflation and
dark energy in the presence of a single perfect fluid\footnote{If two fluids
are present (for example, matter and radiation) simultaneously then an exact
solution requires the scalar field to have a more complicated potential than
the power of an exponential. Our model gives the limiting form of that more
complicated solution in the limiting cases where one perfect fluid dominates
over the other.}. Finally, following the notations of \cite{Peea} we
can provide a unified picture of dark energy and inflation using the sum of
the potentials (\ref{pot}) and (\ref{pot1})
\[
V_{tot}(\phi)=A_{de}\;\sinh^{b_{de}}\left[  \lambda_{de}\left(  \phi-\phi
_{0}\right)  \right]  +A\;\mathrm{sinh}^{b}(\phi/f)
\]
where $\lambda_{de}=\frac{\sqrt{3}(1-\gamma_{\phi})}{\sqrt{\gamma_{\phi}}}$ and the
constants $A_{de}$, $b_{de}$, $b$, are given by Eqs.(\ref{AA}), (\ref{bb}) and
(\ref{bbw}). Note that similar considerations hold for the early dark-energy
model \cite{Bar00}.

\textbf{Acknowledgments.} SB acknowledges support by the Research Center for
Astronomy of the Academy of Athens in the context of the program
\textquotedblleft\textit{Tracing the Cosmic Acceleration}\textquotedblright.
JDB acknowledges STFC support. Also, we would like to thank N. E. Mavromatos
for useful comments and suggestions.


\end{document}